# Using a Performance Model to Implement a Superscalar CVA6

Côme Allart[1,2], Jean-Roch Coulon[1], André Sintzoff[1], Olivier Potin[2], Jean-Baptiste Rigaud[2]

[1] *Thales DIS*, Meyreuil, France

[2] *Mines Saint-Etienne, CEA, Leti, Centre CMP*, F-13541 Gardanne, France

[1] {come.allart, jean-roch.coulon, andre.sintzoff}@thalesgroup.com

[2] {come.allart, olivier.potin, rigaud}@emse.fr

**Abstract**  *A performance model of CVA6 RISC-V processor is built to evaluate performance-related modifications before implementing them in RTL. Its accuracy is 99.2% on CoreMark. This model is used to evaluate a superscalar feature for CVA6. During design phase, the model helped detecting and fixing performance bugs. The superscalar feature resulted in a CVA6 performance improvement of 40% on CoreMark.*

## 1. Introduction

RISC-V is an open Instruction Set Architecture (ISA); organizations can build processors implementing this ISA without fees. CVA6 is an open-source RISC-V processor created by ETH Zurich [1] and maintained by OpenHW Group [2]. It is highly-configurable to address industrial applications of many OpenHW Group members with a single codebase. This single-issue processor issues instructions in-order but executes them out-of-order. CVA6 codebase is a synthesizable Register-Transfer Level (RTL) model.

There is a need to improve CVA6 performance to address more applications. For instance, [3] implemented a superscalar CVA6 and reported a performance gain of 16% on Dhrystone benchmark. To investigate further performance improvements, we propose a model-driven methodology for architectural exploration and implementation.

Modeling only the control path eases modifications for design exploration while allowing performance modeling. Hence, detailed models such as Gem5 [4] are not suitable. Cycle-accurate comparison with RTL is needed for model-driven implementation so mechanistic and empirical modeling [5] are not suitable either. Despite the short *execution* time of such models, we focus on limiting the *development* time to perform microarchitecture exploration.

We implemented a cycle-accurate model of the reference CVA6 in Python. We modified this model to perform architectural exploration and take design decisions. We then performed a model-driven implementation of a superscalar CVA6: we often used our model to verify that the expected performance is met, and to debug potential discrepancies. The *superscalar* feature is optional and can be enabled by RTL configuration.

CoreMark [6] is a well-known benchmark with matrix operations, chained lists and state machines. An iteration runs about 266,000 dynamic instructions. The performance of the reference CVA6 is 3.10 CoreMark/MHz [7] and 4.35 CoreMark/MHz for the superscalar version resulting from our model-driven approach, improving performance by 40%.

Our model is open-source[1] and our superscalar CVA6 will be proposed for integration into CVA6 codebase.

The paper is organized as follows. First, we detail our model. We then describe the main changes to build a superscalar CVA6. Finally, implementation results are presented.

---

[1] See https://github.com/ThalesSiliconSecurity/cva6-perf-model

# 2. Modeling

Our model simulates performance but not behavior. Hence, it requires a RISC-V Formal Interface (RVFI) [8] trace as an input, i.e. the list of dynamic instructions executed. It does not run faster than Verilator [9] RTL simulation. Developed in Python, it is easier to modify as it is 24 times smaller than CVA6 RTL model. CVA6 hardware architecture is a 6-stage pipeline:

1. Instruction fetch
2. Program counter generation (branch prediction)
3. Decode
4. Issue (operands reading)
5. Execute
6. Commit

Only pipeline stages 4 to 6 are modeled because they require most design decisions. These three stages are modeled with Python functions as shown in Figure 1. A clock cycle is modeled by running these three functions. They are run in reverse order to simulate sequential logic [10].

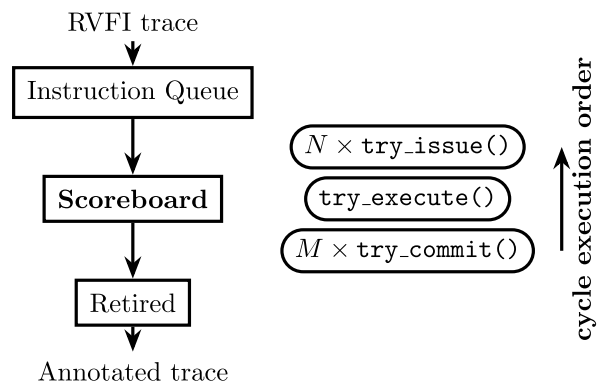

Figure 1: The pipeline of the model

## 2.a. Modeling instruction queue

A list is built from the input RVFI trace where each dynamic instruction is stored as an object with the following fields:

- the program counter, for branch prediction;
- the instruction in hexadecimal, for decoding;
- the disassembly, for human-readable display.

This list, called *Instruction Queue*, is a FIFO because CVA6 is an in-order processor.

## 2.b. Modeling issue stage

Issuing consists in taking an instruction from *Instruction Queue* and pushing it into *Scoreboard*. The latter keeps track of the issued-but-not-committed instructions, as it does in the RTL. `try_issue()` verifies the absence of any of the three kinds of hazards [11] beforehand.

Data hazards between the instruction being issued $i$ and each scoreboard entry $e$ are checked. There are three sorts of data hazards. Read After Write (RAW) hazards occur when one of the source registers of $i$ is the destination register of $e$. CVA6 can forward operands: there is no need to stall if the result from $e$ is available albeit not committed. Write After Write (WAW) hazards happen when the destination register of $i$ is the same as $e$. These hazards can be eliminated thanks to register renaming. CVA6 has no Write After Read (WAR) hazards because it issues instructions in-order.

Structural hazards occur when *Scoreboard* is full and cannot accept new instructions. They are modeled on functional units (FUs) with a `busy` Boolean for each FU, indicating that it cannot accept new instructions. When an instruction is issued to FU $f$, it becomes `busy` because FUs can accept at most one instruction per cycle. If another FU uses the same write-back port as $f$, it becomes `busy` as well. As a new cycle starts, FUs are freed. However, as the multiplier has a 2-stage internal pipeline; if it was `busy` the previous cycle, units sharing its write-back port become `busy` instead.

Control hazards are handled by performing branch prediction on the last issued instruction. In case of a branch miss, issue stage stalls if the miss occurred less than 6 cycles ago. CVA6 has three branch prediction mechanisms: Return Address Stack (RAS), Branch History Table (BHT), and Branch Table Buffer (BTB). RAS and BHT model implementations are similar to their CVA6 counterparts. BTB is not part of the model because there are no indirect jumps in CoreMark; except returns, which are handled by the RAS.

The corresponding `try_issue()` function is called $N$ times per cycle to model a hypothetical $N$-issue CVA6.

### 2.c. Modeling execute stage

Each instruction is issued into *Scoreboard* with a cycle counter initialized to 0. `try_execute()` increments all these counters every cycle because issued instructions are executed in parallel. When a counter reaches the instruction execution duration, its instruction is marked as *done*. Multiplications, loads and stores are 2-cycle long whereas other instructions take 1 cycle. This `try_execute()` function is called once per clock cycle.

### 2.d. Modeling commit stage

*done* instructions are removed from *Scoreboard* in-order and put into the *Retired* list. This `try_commit()` function is called $M$ times per cycle to model a hypothetical $M$-commit CVA6 but only the first commit port can commit a store.

### 2.e. Measuring model accuracy

The model outputs an RVFI trace annotated with the commit cycle of each dynamic instruction. To validate our model, we need to compare this trace with one from the RTL. We modified the RVFI tracer in RTL to get commit cycle from Verilator simulation. To measure accuracy, we used the methodology from [7], which we recall below.

Let $t_i$ be the commit cycle of the $i^{\text{th}}$ instruction. Let $\Delta t_i = t_i - t_{i-1}$. For each instruction $i$, this duration from the model $\Delta t_i^{\text{Model}}$ is compared with the one from Verilator $\Delta t_i^{\text{RTL}}$. We define the accuracy as the ratio of matching instructions.

$$\text{Accuracy} = \frac{\#\{i \mid \Delta t_i^{\text{Model}} = \Delta t_i^{\text{RTL}}\}}{\#\{i\}}$$

We measured an accuracy of 99.2% on CoreMark $2^{\text{nd}}$ iteration. We used a `cv32a6_imac_sv0`-based CVA6 configuration, enabling Zba, Zbb, Zbc and Zbs RISC-V extensions support and assuming a low cache latency of one cycle. When disabling the RAS in both the model and the RTL, the model accuracy increases to 99.8%. From this promising accuracy, we will exploit it to take implementation decisions.

### 2.f. First results from the model

[7] focused on issue width $I$ and the number of commit ports $C$. They report no benefit to have $C > I$ and little benefit to have $I > C$ while increasing $I$ requires more hazards management logic so we chose $I = C$. Considering performance and pipeline widths, we chose a width of 2.

The only structural hazard modeled is the shared write-back port between the ALU and the multiplier, which have different latencies. It is the only structural hazard that occurs on CoreMark in single-issue CVA6. We improved the modeling of structural hazards as explained in Section 2.b. A second ALU was added with precedence for better FU usage.

We discarded modifications which were predicted to not bring enough performance according to the model. For instance, we initially planned to connect the multiplier to the same write-back port as the FPU instead of the ALU. The hypothesis was that integer multiplications are rarely mixed with floating-point operations. However, they are often mixed with integer additions, for instance to multiply-accumulate. As the model predicted a performance gain of less than 3% for a single-issue CVA6 on the CoreMark, we decided to discard this modification.

## 3. Implementing

To implement a 2-way superscalar CVA6, we split our development in three steps. Expected performance improvements below are relative to single-issue CVA6. The implementation step definition is comparable to [3], minus branch prediction.

- 64-bit instruction fetch: +1%
- Dual issue (single ALU): +21%
- Superscalar (two ALUs): +47%[2]

Our model provides 3 levels of comparison with RTL results from Verilator for debugging.

($\alpha$) Global performance: cycles spent.
($\beta$) Local performance: cycle-annotated trace.
($\gamma$) Internal: print scoreboard and events each cycle.

We evaluate performance ($\alpha$) after each development step. If we are satisfied, we continue with the next step. Else, we compare the annotated trace ($\beta$) from Verilator with the one from our model to find which instruction sequences cause the difference. The issue can be either in our RTL implementation or in the model. To understand the difference, we compare the cycle-accurate scoreboard evolution ($\gamma$) using Verilator waveforms and the model debug trace.

The modified parts of CVA6 are shown in Figure 2, where colors indicate three levels of changes. Blue parts are affected by parameter value changes without significant code changes. Purple parts are modified for genericity or new features. Green modules are instantiated more times in dual- than in single-issue. These modifications can be enabled via SystemVerilog genericity features.

### 3.a. Frontend

CVA6 *Frontend* gathers stages 1 and 2. Making instruction fetch from 32- to 64-bit required few changes as it was almost generic. We fixed *Re-aligner* 64-bit support.

*Instruction Scan* detects control flows for branch prediction. Instructions are put into *Instruction Queue* before the decode stage. CVA6 conditionally supports compressed instructions so these items were already generic.

### 3.b. Dual issue

We made it possible to pop up to two instructions from *Instruction Queue*. We duplicated *Compressed Decoder* and *Decoder*. We made *Issue Buffer* a 2-instruction FIFO.

---

[2]Like the reference CVA6, the model had register renaming at the time of the experiments. This feature has since been removed from CVA6. The model has been updated since, removing register renaming and integrating feedback from the implementation phase. The improvement predicted by the latest version of the model is 41%.

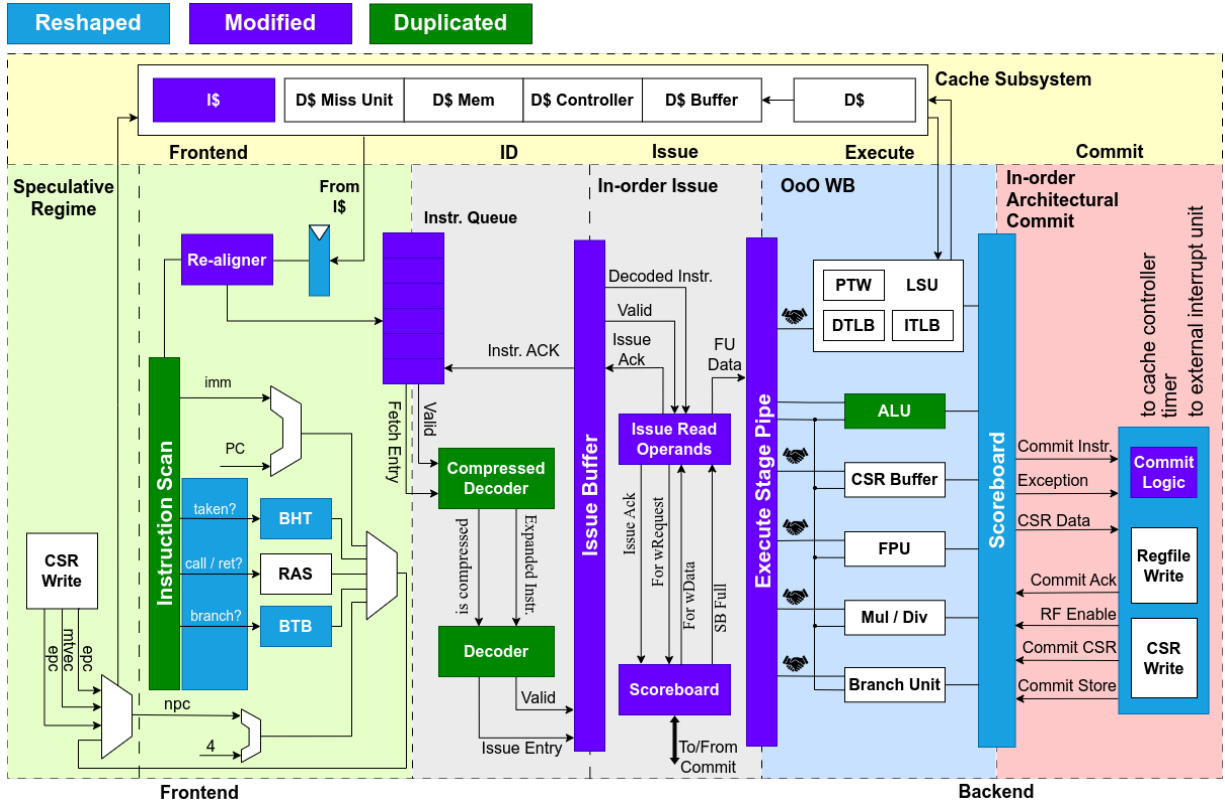

Figure 2: Simplified version of CVA6 schematic [2], with colors for superscalar-related modifications.

*Issue Read Operands* reads twice more operands from registers and detects twice more data hazards with instructions in *Scoreboard*, allowing twice more operand forwarding. It detects hazards between the two instructions being issued.

We implemented structural hazards detection in the same way as we did in our Python model in Section 2.b.

After this step, the superscalar made the embedded version of CVA6 slower. We found with the model ($\gamma$) that *Scoreboard* was not efficiently used anymore, which became an issue due to embedded little *Scoreboard*. Indeed, we initially prevented from issuing as soon as less than two *Scoreboard* entries were free. We fixed this by detecting separately if *Scoreboard* is full and if there is exactly one free entry. To do this, we check if all odd index entries are occupied. We do the same for even index entries. As *Scoreboard* is a circular buffer, an *and* between these two values means that the scoreboard is full and an *or* means that a second instruction cannot be issued this cycle.

*Issue* stage now emits two instructions to *Execute* stage. We added a multiplexer 2–1 before each FU. It selects an issue port targeting this FU. No precedence rule is required because *Issue* stage already handles structural hazards.

### 3.c. Speculative Scoreboard

At the end of the dual-issue implementation step, a performance improvement of 12% was reached, instead of 21%. The cycle-annotated traces ($\beta$) showed that most differences followed branch hits. Indeed, in the model, we did not implement control hazards between instructions issued in the same cycle. To fill the gap, the instruction following a control flow should be issued the same cycle. Consequently, it must be discarded when the branch is resolved as missed. However, *Scoreboard* could only be completely flushed.

*Scoreboard* is a FIFO with two pointers: the issue pointer to push instructions and the commit one to pop them. We wrote a new module to build the interval of instructions to discard, which are between the branch and the issue pointer. It takes two indexes $(A, B) \in [0; N[^2$ and returns a $N$-bit vector where the bit at index $i$ is set if $i \in [A; B[$. *Scoreboard* is cycling, so $A > B$ is valid and results in a non-zero vector.

To discard instructions, simply removing them from *Scoreboard* does not work because the removed instruction is not cancelled in *Execute* stage: it can write back the wrong result into the entry of the correct instruction. Instead, a `cancelled` bit was added to each *Scoreboard* entry. When a branch is missed, the `cancelled` bits of next instructions are set. *Commit* stage acknowledges cancelled instructions without modifying the architectural state.

This *speculative scoreboard* feature is optional.

### 3.d. Second ALU

To not add an area-consuming write-back port, the added ALU uses the *FPU* write-back port. However, knowing when the FPU will yield a result is not trivial. As we do not need a FPU yet, we chose to not spend time analyzing it. As a consequence, superscalar CVA6 does not support FPU yet.

## 4. Results

Performance, Power and Area (PPA) results are shown in Table 1. The reference is the same configuration as in Section 2.e. For superscalar we enabled the *superscalar* and *speculative scoreboard* options. It results in a performance of 4.35 CoreMark/MHz; a gain of 40% which is greater than the related 11% area cost. The maximum frequency slightly reduced but the critical path is not in the modified parts.

| Criteria | Reference | Superscalar | Variation |
|---|---|---|---|
| CoreMark/MHz | 3.10 | **4.35** | **+40.1%** |
| Max. Frequency | 892 MHz | 877 MHz | −1.75% |
| Power | 32.45 mW | 34.84 mW | +7.37% |
| Area | 250 kGE | 278 kGE | +11.1% |

Table 1: Performance / Power / Area results

All basic tests from the CVA6 repository pass with and without *superscalar* enabled. Linux boots on FPGA for single-issue and *superscalar* without *speculative scoreboard*. Booting Linux with *speculative scoreboard* is left for future work.

We also ran Dhrystone on our superscalar CVA6 and measured a 24% gain; 1.5 times the previous version gain. This tends to validate our model-driven approach.

## 5. Discussion

Our work is based on a model that is accurate and easy to modify. As a consequence, the model ignores data. It prevents the modeling of divisions and data cache. By extracting data from the RVFI trace, we could model them. Also, speculative instructions are not present in the RVFI trace. It prevents the modeling of instruction cache. Still, the model accuracy on CoreMark $2^{nd}$ iteration is above 99% so we could take decisions based on it. However, as we focused on the CoreMark, the model accuracy could be lower on other benchmarks.

We noticed that CoreMark has few WAW hazards compared to Dhrystone. By activating register renaming in our model, the gain related to the *superscalar* feature reaches 45% on CoreMark. Thus, we believe that register renaming could significantly improve CVA6 performance on Dhrystone.

# 6. Conclusion

A performance model of CVA6 was built in raw Python. An accuracy metric was defined. The measured model accuracy is above 99%. We extrapolated the model to estimate the performance gain of making CVA6 superscalar.

A superscalar CVA6 was implemented and can be enabled by configuration. While implementing, the model was used as a reference with means of detailed comparison. It helped finding and fixing several performance bugs. Especially, we added an optional *speculative scoreboard* feature which allows partial scoreboard flush.

Future works will focus on (i) booting Linux with this new feature and pushing these changes to the CVA6 repository [2]; (ii) implementing register renaming via the scoreboard to fix WAW hazards; (iii) running Post Quantum Cryptography (PQC) benchmarks using the model to see how it matches RTL and to evaluate potential optimizations.

# Acknowledgments

These activities are supported by the TRISTAN project funded by the Key Digital Technologies Joint Undertaking (KDT JU) under grant agreements 101095947. The present action reflects only the authors' view; the European Commission and the JU are not responsible for any use that may be made of the information it contains.